\documentclass[floatfix,showpacs]{revtex4}
\usepackage{epsfig}
\usepackage{graphicx}
\newcommand{\vphi}{\varphi}
\newcounter{fig}   \newcommand{\lbfig}[1]{\refstepcounter{fig}
\label{#1} } 

\begin{document}

\title{Gravitating Dyons with Large Electric Charge}

\vspace{1.5truecm}
\author{Rustam Ibadov}
\affiliation{Department of Theoretical Physics and Computer Science,\\
Samarkand State University, Samarkand, Uzbekistan}
\author{Burkhard Kleihaus}
\affiliation{ZARM, Universit\"at Bremen, Am Fallturm,
D--28359 Bremen, Germany}
\author{Jutta Kunz and Ulrike Neemann}
\affiliation{Institut f\"ur Physik, Universit\"at Oldenburg, Postfach 2503,
D--26111 Oldenburg, Germany}

%\pacs{04.20.Jb, 04.40.Nr}  
\date{\today}

\begin{abstract}
We consider non-Abelian dyons in Einstein-Yang-Mills-Higgs theory.
The dyons are spherically symmetric with unit magnetic charge.
For large values of the electric charge the dyons approach limiting solutions, 
related to the Penney solutions of Einstein-Maxwell-scalar theory.
\end{abstract}
\maketitle

\section{Introduction}

Magnetic monopoles arise as regular non-perturbative solutions
of $SU(2)$ Yang-Mills-Higgs theory \cite{mono}.
Monopoles with magnetic charge $n=1$ are spherically symmetric,
monopoles with higher charge have axial symmetry
\cite{WeinbergGuth,RebbiRossi,mmono,KKT}
or no rotational symmetry at all \cite{monoDS}.
When electric charge is added to magnetic monopoles,
dyons arise \cite{dyon,dyonhkk}. 
These are stationary solutions with vanishing
angular momentum \cite{HSV,radu}.

The nontrivial vacuum structure of $SU(2)$ Yang-Mills-Higgs theory
not only allows for magnetic monopoles, but it results in
a plethora of further regular non-perturbative solutions
\cite{Taubes,Rueber,mapKK,KKS,KNS}.
The simplest of these solutions are 
monopole-antimonopole pairs,
where a monopole and an antimonopole % located on the symmetry axis
form an unstable equilibrium configuration
\cite{Taubes,Rueber,mapKK}.

When gravity is coupled to Yang-Mills-Higgs theory,
gravitating monopoles \cite{gmono,HKK}, dyons \cite{gdyon},
monopole-antimonopole pairs \cite{MAP,PRT,KKN,KKNN}, 
% charged rotating MAPs \cite{PRT,KKN,KKNN},
and further configurations arise \cite{KKS4}.
These solutions depend on a dimensionless coupling constant $\alpha$,
which is proportional to the square root of Newton's constant % $G$
and the Higgs vacuum expectation value. % $v$.

For each type of solution, a branch of gravitating solutions
emerges smoothly from the corresponding flat space solution
and extends up to a maximal value of $\alpha$,
beyond which the size of the core of the solution would be smaller than
its Schwarzschild radius \cite{gmono}.
At the maximal value of $\alpha$ this fundamental branch
bifurcates with a second branch of solutions.

For monopoles and dyons,
in the case of vanishing Higgs potential, 
this second branch reaches a critical value of $\alpha$,
where it bifurcates with
the branch of extremal Reissner-Nordstr\"om black holes
with the same charge(s) \cite{gmono,HKK,dyon}.
For monopole-antimonopole pair solutions
and other composite solutions, in contrast,
the second branch extends back to $\alpha=0$,
%where the solutions shrink to zero size.
where a pure Einstein-Yang-Mills solution is reached
(after scaling w.r.t.~$\alpha$) \cite{BM,KK,IKKS,Kari}.

As the second monopole resp.~dyon branch reaches the
critical value of the coupling constant $\alpha$,
and merges with the extremal Reissner-Nordstr\"om branch,
%an infinitely long throat forms, 
the spacetime splits into two regions \cite{gmono}.
The exterior spacetime of the critical solution, 
corresponds to the one of an extremal Reissner-Nordstr\"om
black hole \cite{gmono},
while the interior spacetime retains regularity at the center,
due to the influence of the non-Abelian fields present.

Here we reconsider gravitating dyons and study
the properties to the solutions in
the limit of large electric charge $Q$.
We demonstrate, that for $Q \rightarrow \infty$
we also obtain a limiting solution,
which consists of two regions, 
a non-Abelian interior regions 
and an Abelian exterior region.
But in this case, the non-Abelian interior part has a flat metric,
and the Abelian exterior part corresponds to
a certain Penney solution of Einstein-Maxwell-scalar theory
(after scaling w.r.t.~$Q$) \cite{Penney:1969xk}.

In section II we present the 
Einstein-Yang-Mills-Higgs action, the Ansatz, 
the equations of motion and the boundary conditions.
We then consider the equations for large electric charge,
and relate them to the equations of Einstein-Maxwell-scalar theory.
We describe the relevant features of the Penney solutions in section III,
and discuss the properties of the numerically constructed
non-Abelian dyons in section IV,
where we also address their relation to the Penney solutions.
We present our conclusions in section V.

\section{\bf Einstein-Yang-Mills-Higgs Solutions}

\subsection{Action}

We consider the SU(2) Einstein-Yang-Mills-Higgs action
in the limit of vanishing Higgs potential,
\begin{eqnarray}
S &=& \int \left [ \frac{R}{16\pi G}
  - \frac{1}{2} {\rm Tr} \left(F_{\mu\nu} F^{\mu\nu}\right)
 -\frac{1}{4} {\rm Tr} \left( D_\mu \Phi D^\mu \Phi \right)
 \right ] \sqrt{-g}\, d^4x
\ \label{action} \end{eqnarray}
with curvature scalar $R$,
SU(2) field strength tensor
\begin{equation}
F_{\mu \nu} =
\partial_\mu A_\nu -\partial_\nu A_\mu + i e \left[A_\mu , A_\nu \right]
\ , \label{fmn} \end{equation}
gauge potential $A_\mu = 1/2 \tau^a A_\mu^a$,
gauge covariant derivative
\begin{equation}
D_\mu = \nabla_\mu +ie [ A_\mu, \cdot \ ]
\ , \label{Dmu} \end{equation}
and Higgs field $\Phi = \tau^a \Phi^a$;
$G$ is Newton's constant, and $e$ is the gauge coupling constant.
Since we here consider vanishing Higgs potential,
we impose a Higgs field vacuum expectation value $v$.

Variation of the action Eq.~(\ref{action}) with respect to the metric
$g_{\mu\nu}$, the gauge potential $A_\mu^a$, and the Higgs field $\Phi^a$
leads to the Einstein equations and the matter field equations,
respectively.

\subsection{Ans\"atze}

We employ Schwarzschild-like coordinates 
and parametrize the line element by \cite{gmono,gdyon}
\begin{equation}
ds^2 = -A(r)^2 N(r)\, dt^2 + \frac{1}{N(r)}\, dr^2 + r^2 d \Omega^2 \, ,
\ \ \ N(r)=1-\frac{2 \mu(r)}{r} \, ,
\end{equation}
with mass function $\mu(r)$.
The Ansatz for the gauge potential and Higgs field is given by
\begin{equation}
A_\mu dx^\mu=
\frac{1-H_2(r)}{2e} \left( \tau_\varphi d\theta
                        -\tau_\theta \sin\theta d\varphi \right)
+\frac{B_1(r)}{2e} \tau_r dt
\ , \ \ \
\Phi = v \Phi_1(r) \tau_r \ ,
\end{equation}
where the $su(2)$ matrices
$\tau_r$, $\tau_\theta$, and $\tau_\vphi$
are defined as scalar products of the spatial unit vectors
with the Pauli matrices $\tau^a = (\tau_x, \tau_y, \tau_z)$,
and the subscripts on the functions
indicate the correspondence to the functions
of the more general Ansatz, necessary for 
the construction of monopole-antimonopole systems \cite{KK,KKS}.

\subsection{Equations of Motion}

We now change to dimensionless quantities,
the dimensionless coordinate $\tilde r$,
the dimensionless electric function $\tilde B_1$,
and the dimensionless mass function $\tilde \mu$,
\begin{equation}
\tilde r = e v r \ , \ \ \ \tilde B_1 = \frac{B_1}{e v}
                 \ , \ \ \ \tilde \mu=e v \mu
\ , \label{xm} \end{equation}
and introduce the dimensionless coupling constant $\alpha$
\begin{equation}
\alpha^2 = 4 \pi G v^2 
\ . \end{equation}

Suppressing the $\tilde {\phantom{r}}$ in the following,
the $tt$ and $rr$ components of the Einstein equations yield
for the metric functions the equations,
\begin{eqnarray}
\mu'&=&\alpha^2 \Biggl( \frac{r^2 B_1'^2}{2 A^2}
   + \frac{B_1^2 H_2^2}{A^2 N}
   + N H_2'^2 + \frac{1}{2} N r^2 \Phi_1'^2
   + \frac{(H_2^2-1)^2}{2 r^2} + \Phi_1^2 H_2^2
%   + \frac{\beta^2}{4} r^2 (H^2-1)^2 
\Biggr)
\ , \end{eqnarray}
and
\begin{eqnarray}
 A'&=&\alpha^2 r \Biggl(
    \frac{2 B_1^2 H_2^2}{ A^2 N^2 r^2}
   + \frac{2 K'^2}{r^2} + \Phi_1'^2 \Biggr) A
\ , \label{eqa} \end{eqnarray}
while the matter field equations yield
\begin{eqnarray}
(A N H_2')' = A H_2 \left( \frac{H_2^2-1}{r^2} + \Phi_1^2
 - \frac{B_1^2}{A^2 N}
 \right)
\ , \end{eqnarray}
\begin{equation}
\left( \frac{r^2 B_1'}{A} \right)' = \frac{2 B_1 H_2^2}{AN}
\ , \end{equation}
and
\begin{eqnarray}
( r^2 A N \Phi_1')' = 2 A \Phi_1  H_2^2 
%+ \beta^2 r^2 (H^2-1) 
\ . \end{eqnarray}
These equations then
depend only on the dimensionless coupling constant $\alpha$ \cite{gmono}.

A particular solution of these equations is the embedded 
Reissner-Nordstr\"om solution
with unit magnetic charge, $P=1$, and arbitrary electric charge $Q$,
\begin{equation}
\mu(r) = m - \frac{\alpha^2 (Q^2+1)}{2r} , \ \ \ A(r)=1
\ , \end{equation}
\begin{equation}
H_2(r)=0 \ , \ \ \
B_1(r) = \nu - \frac{Q}{r} \ , \ \ \
\Phi_1(r)=1
\ , \label{Q} \end{equation}
where the extremal solution satisfies
\begin{equation}
m = r_{\rm H} = \alpha \sqrt{Q^2+1}
\ . \label{extr} \end{equation}

\subsection{Boundary Conditions}

Dyons are globally regular particle-like solutions
of the SU(2) Einstein-Yang-Mills-Higgs system.
Regularity of the solutions at the origin then requires
the boundary conditions
\begin{equation}
\mu(0)=0
\ , \end{equation}
and \cite{dyon}
\begin{equation}
H_2(0)=1 \ , \ \ \ B_1(0) = 0 \ , \ \ \ \Phi_1(0)=0
\ . \end{equation}

Asymptotic flatness of the solutions, on the other hand, implies
that the metric functions $A$ and $\mu$ both
approach constants at infinity. We adopt
\begin{equation}
A(\infty)=1
\ . \end{equation}
The matter functions satisfy the asymptotic boundary conditions
\begin{equation}
H_2(\infty)=0 \ , \ \ \ B_1(\infty)= \nu \ , \ \ \ \Phi_1(\infty) = 1
\ . \end{equation}

The dimensionless magnetic charge $P$ is the topological charge
of the solutions, 
while the dimensionless electric charge $Q$,
and the dimensionless scalar charge $c_{\rm H}$ are obtained from
the asymptotic expansion of the fields, thus
\begin{equation}
P = 1
\ , \ \ \
Q = -\lim_{r \rightarrow \infty} r \left( B_1- \nu \right)
\ , \ \ \
c_{\rm H} =   \lim_{r \rightarrow \infty} r^2 \partial_r \Phi_1
\ , \label{QP} \end{equation}
and $\mu(\infty)=m$ represents the dimensionless mass
of the solutions.

\subsection{Scaled equations of motion}

Since we are interested in the limit of large electric charge,
$Q \rightarrow \infty$, we now consider the above set of equations
in terms of quantities scaled by the charge $Q$. 
Thus we introduce
\begin{equation}
\bar r = \frac{r}{Q} \ , \ \ \
\bar m =\frac{m}{Q} \ , \ \ \
\bar c_{\rm H} = \frac{c_{\rm H}}{Q} \ .
\label{bar} 
\end{equation}

In the limit $Q \rightarrow \infty$ we then obtain
a coupled set of equations for $\mu$, $A$, $B_1$ and $\Phi_1$,
\begin{equation}
\bar \mu ' =
\frac{\alpha^2}{2\ } \bar r^2 \left( \frac{ B_1'^2}{A^2}
 + N \Phi_1'^2 \right) \ , 
\label{eqmu} \end{equation}
\begin{equation}
 A' = \alpha^2 \bar r \Phi_1'^2 A \ , 
\label{eqA} \end{equation}
\begin{equation}
\frac{ \bar r^2 B_1'}{A} = \bar Q = 1 \ , 
\label{eqB1} \end{equation}
\begin{equation}
\bar r^2 A N \Phi_1' = \bar c_{\rm H} \ ,
\label{eqP1} \end{equation}
where the $ {'}$ denotes differentiation w.r.t.~$\bar r$,
while the equation for $H_2$ decouples,
\begin{equation}
H_2 {''}= \frac{H_2(H_2^2-1)}{\bar r^2} \ .
\label{eqH2} \end{equation}
The coupled set of equations Eqs.~(\ref{eqmu})-(\ref{eqP1})
corresponds precisely
to the set of equations of Einstein-Maxwell-scalar theory,
when the electric charge has the value $Q=1$.
We therefore now turn to the Einstein-Maxwell-Scalar Solutions.

\section{\bf Einstein-Maxwell-Scalar Solutions}

%We briefly review the Penney solution \cite{Penney:1969xk},
%and subsequently specialize to the case associated with the
%limiting exterior dyon solution.

\subsection{Action}

The matter Lagrangian of Einstein-Maxwell-Scalar theory reads
\begin{equation}
{\cal L} = - \frac{1}{4} F_{\mu\nu}F^{\mu\nu} 
 - \frac{1}{2} \partial_\mu \Phi_1 \partial^\mu \Phi_1 
\end{equation}
with the Abelian gauge potential $A_\mu d x^\mu$ 
and the scalar field $\Phi_1$,
giving rise to the stress-energy tensor in the 
Einstein equations,
\begin{equation}
G_{\mu\nu}= 2 \alpha^2 T_{\mu\nu} \ ,
\end{equation}
in terms of dimensionless coordinates and fields.
%(Penney denotes $2 \alpha^2=K$).

\subsection{Penney solutions}

Penney obtained static spherically symmetric solutions
with electric charge $Q$ % (Penney's $\epsilon$)
and scalar charge $c_{\rm H}$. % (Penney's $c$).
He employed the metric parametrization
\begin{equation}
ds^2 = - e^{-a} dt^2 + e^{a} dR^2 + e^{b}
 d \Omega^2 \ , 
\end{equation}
with radial coordinate $R$. 

The Penney solutions then read
\begin{equation}
e^{b} = \left( (R-\rho)(R-\sigma) \right)^{-\Lambda}
 \left( \frac{ \sigma(R-\rho)^{\Lambda} - \rho(R-\sigma)^{\Lambda} }
             {\sigma-\rho} \right)^2
\ , \ \ \
e^{b} = \left( (R-\rho)(R-\sigma) \right) e^{a} \ ,
\end{equation}
\begin{equation}
e^{b} \hat B_1' = Q \ , \ \ \
e^{b - a} \Phi_1' = c_{\rm H} \ ,
\label{eqsP}
\end{equation}
where $A_0=\hat B_1$ denotes the time component
of the gauge potential.

The constants $\rho$, $\sigma$, and $\Lambda$ of
the Penney solutions satisfy the relations
\begin{equation}
\Lambda^2 \rho \sigma = \alpha^2 Q^2 \ , \ \ \
 \alpha^2 c_{\rm H}^2 = \left( 1- \Lambda^2 \right)
  \left( \frac{\rho-\sigma}{2} \right)^2 \ .
\label{rel} \end{equation}
%and consequently $\Lambda^2 \le 1$.

Since we want to relate the Penney solutions to
the limiting dyon solutions, we must select 
those Penney solutions, which do not exhibit a metric singularity
anywhere except the origin.
Thus we now focus on the case,
where in the Reissner-Nordstr\"om limit
no horizons occur, but a naked singularity is present.
%We now consider solutions with $g_{tt}>0$.
In the Penney solutions
this is achieved by the choice $\rho=\rho_0+i \sigma_0$, 
$\sigma=\rho_0-i \sigma_0$.
Defining
\begin{equation}
\xi^2 = (R-\rho_0)^2 + \sigma_0^2 \ , \ \ \
\tan \Psi = \frac{\sigma_0}{R-\rho_0} \ , \ \ \
\tan \varphi_0 = -\frac{\rho_0}{\sigma_0} \ , \ \ \
\gamma^2 = \rho_0^2 + \sigma_0^2 \ , 
\end{equation}
then yields the metric functions
\begin{equation}
e^{a} = \left( \frac{ \cos(\Lambda \Psi - \varphi_0)}
 {\cos \varphi_0} \right)^2  \ , \ \ \
e^{b} = \xi^2 e^{a}
\label{metP} \end{equation}
and the relations, Eq.~(\ref{rel}),
\begin{equation}
\Lambda^2 = \frac{\alpha^2 Q^2}{\gamma^2} \ , \ \ \
 \cos^2 \varphi_0 =
 \frac{\alpha^2 c_{\rm H}^2} { \alpha^2 Q^2- \gamma^2} \ , \ \ \
 m^2 = \alpha^2 Q^2 \sin^2 \varphi_0 \ ,
\label{rel2} \end{equation}
where the mass $m$ is obtained from the asymptotics.
From $\sin^2 \varphi_0 + \cos^2 \varphi_0=1$ we obtain
\begin{equation}
\frac{1}{\Lambda^2 } = 1 - \frac{c_{\rm H}^2}{Q^2 - \left(
 \frac{m}{\alpha} \right)^2 }  \ . %\ \ \
\label{rel3a} \end{equation}
%\Longrightarrow \ \ \
Since $\Lambda^2 \ge 0$ we obtain the bound
\begin{equation}
c_{\rm H}^2 \le Q^2 - \left( \frac{m}{\alpha} \right)^2 \ .
\label{rel3} \end{equation}

\subsection{Limit of large electric charge}

We now consider the limit of large charge, taking $Q \rightarrow \infty$. 
Then according to Eq.~(\ref{rel2}) also $\Lambda \rightarrow \infty$. 
Eq.~(\ref{rel3a}) then requires that
the bound in Eq.~(\ref{rel3}) is precisely saturated, 
\begin{equation}
c_{\rm H}^2 = Q^2 - \left( \frac{m}{\alpha} \right)^2 \ .
\label{rel4} \end{equation}

To obtain the limiting Penney solution for $Q \rightarrow \infty$,
we thus impose this bound. We then consider the set of equations
w.r.t.~the scaled coordinate $\bar R= R/Q$,
and introduce the scaled quantities
$\bar Q= Q/Q=1$, $\bar \gamma= \gamma/Q$, and $\bar c_{\rm H} = c_{\rm H}/Q$.
%scaled by the charge $Q$.
In particular, we reexpress the relation
$\alpha^2 Q^2 = \gamma^2 \Lambda^2 $ as
$\alpha^2 \bar Q^2 = \bar \gamma^2 \Lambda^2 $,
i.e., $\bar \gamma \rightarrow 0$.
In this limit
\begin{equation}
\cos^2 \varphi_0 =  \bar c_{\rm H}^2 % \frac{c_{\rm H}^2}{Q^2} \ , \ \ \
\ , \ \ \
e^{a} =
%\left( \frac{ \cos(\Lambda \Psi - \varphi_0)} {\cos \varphi_0} \right)^2 = 
%\left( \cos \left( \frac{\alpha c_{\rm H}}{R} - \varphi_0 \right) 
% \frac{Q}{c_{\rm H}} \right)^2 =
  \left( \cos \left(\frac{\alpha \bar c_{\rm H}}{\bar R} \right)
  + \sqrt{ \frac{1}{\bar c_{\rm H}^2} - 1 } \
         \sin \left(\frac{\alpha \bar c_{\rm H}}{\bar R} \right)   \right)^2
\ , \ \ \
\xi^2 = Q^2 \bar R^2 \ .
\label{rel5} \end{equation}

We are interested in the outer extremum 
of the metric function $e^{a}$,
since this is a particular point of the spacetime.
It will later be identified as the transition point,
where the spacetime of the dyons will split into
a non-Abelian interior region and an Abelian exterior region.

The outer extremum of $e^{a}$ occurs at $\bar R_0$, where
\begin{equation}
\frac{\alpha \bar c_{\rm H}^{\phantom{2}}}{\bar R_0} 
 = \arctan \sqrt{ \frac{1}{\bar c_{\rm H}^2} - 1 } \ , \ \ \
e^{a(\bar R_0)} = \frac{1}{\bar c_{\rm H}^2} \ .
\label{rel6} \end{equation}

We now impose the dyon boundary conditions
on the Maxwell and scalar function,
\begin{equation}
\hat B_1(\bar R_0)=0 \ , \ \ \ \hat B_1(\infty)=\nu \ , \ \ \
  \Phi_1(\bar R_0)=0 \ , \ \ \   \Phi_1(\infty)=1 \ .
\label{bcbc} \end{equation}
Integration of the scalar field Eq.~(\ref{eqsP}) then yields 
\begin{equation}
\Phi_1(\infty) = \bar c_{\rm H}^{\phantom{2}} 
 \int_{\bar R_0}^\infty \frac{1}{\bar R^2} d \bar R 
 = \frac{\bar c_{\rm H}^{\phantom{2}}}{\bar R_0} 
 = 1 \ ,
\label{rel7} \end{equation}
i.e., $\bar R_0 = \bar c_{\rm H}^{\phantom{2}}$,
which together with Eq.~(\ref{rel6}) leads to
\begin{equation}
\cos \alpha = \bar c_{\rm H} \ , % \frac{ c_{\rm H}^{\phantom{2}} }{Q} \ .
\label{rel8} \end{equation}
relating the scaled scalar charge to the coupling constant $\alpha$.

To integrate the gauge field Eq.~(\ref{eqsP}),
we first reexpress the metric function $e^{a}$ via
\begin{equation}
e^{a} = \left( \frac{ \cos \left[ \alpha \left(
 \frac{\bar c_{\rm H}^{\phantom{2}}}{\ \ {\bar R}^{\phantom{Z^Z}}} - 1 \right)
 \right] }
 {\cos \alpha} \right)^2 \ .
\label{rel8a} \end{equation}
Integration then yields a relation between 
the asymptotic value of the gauge potential
and the coupling constant $\alpha$,
\begin{equation}
\hat B_1(\infty) = 
   \int_{\bar R_0}^\infty e^{-b} d \bar R 
= \frac{\sin \alpha}{\alpha} = \nu \ .
\label{rel9} \end{equation}

Addressing finally the transformation to Schwarzschild-like
coordinates, we note that
\begin{equation}
A^2 N = e^{-a} \ , \ \ \
r^2 = e^{b} \ , \ \ \
dR = A dr \ .
\label{rel10} \end{equation}
The transition point $r_0$ is thus given by
$r_0^2 = e^{b(R_0)} = R_0^2 Q^2/c_{\rm H}^2 = Q^2$,
i.e., $\bar r_0=1$.

\section{\bf Results}

%We here discuss the properties of stationary dyons.
%For large values of the electric charge,
%we relate the dyons to the Penney
%solutions of Einstein-Maxwell-scalar theory.

\subsection{Gravitating dyons}

Gravitating dyons have been considered before \cite{gdyon,KKNN}.
We here reconsider gravitating dyons and address 
their dependence on the electric charge $Q$.
We focus on large values of the charge
and, in particular, the limit $Q \rightarrow \infty$.

For dyon solutions in flat space, 
the magnitude of the electric charge is 
uniquely determined by the asymptotic value $\nu$ 
of the electric component of the gauge potential.
This potential parameter $\nu$ is bounded,
$0 \le \nu \le 1$,
and in the limit of vanishing Higgs potential,
one has the monotonic relation between $Q$ and $\nu$,
\begin{equation}
Q(\nu) = \frac{\nu}{\sqrt{1-\nu^2}} \, .
\end{equation}
Thus the charge diverges in the limit $\nu \rightarrow 1$.

For gravitating dyons it was also expected,
that (in the limit of vanishing Higgs potential)
the bound $\nu=1$ would be reached monotonically
and that it would correspond to infinite electric charge \cite{gmono}.

When constructing the dyon solutions numerically,
however, a surprise is encountered.
For fixed coupling constant $\alpha$,
the gravitating dyons do not vary monotonically with $\nu$.
Instead at a maximal value $\nu_{\rm max}$ 
a bifurcation is encountered.
Here a second branch extends slightly backwards towards smaller values
of $\nu$.
These bifurcating branches are exhibited in Fig.~\ref{f-1},
where we demonstrate the dependence of the charge $Q$
and the scaled mass $\bar m=m/Q$ on the parameter $\nu$.

\begin{figure}[t!]
\lbfig{f-1}
\begin{center}
\hspace{-0.5cm} (a)\hspace{-0.3cm}
\includegraphics[height=.25\textheight, angle =0]{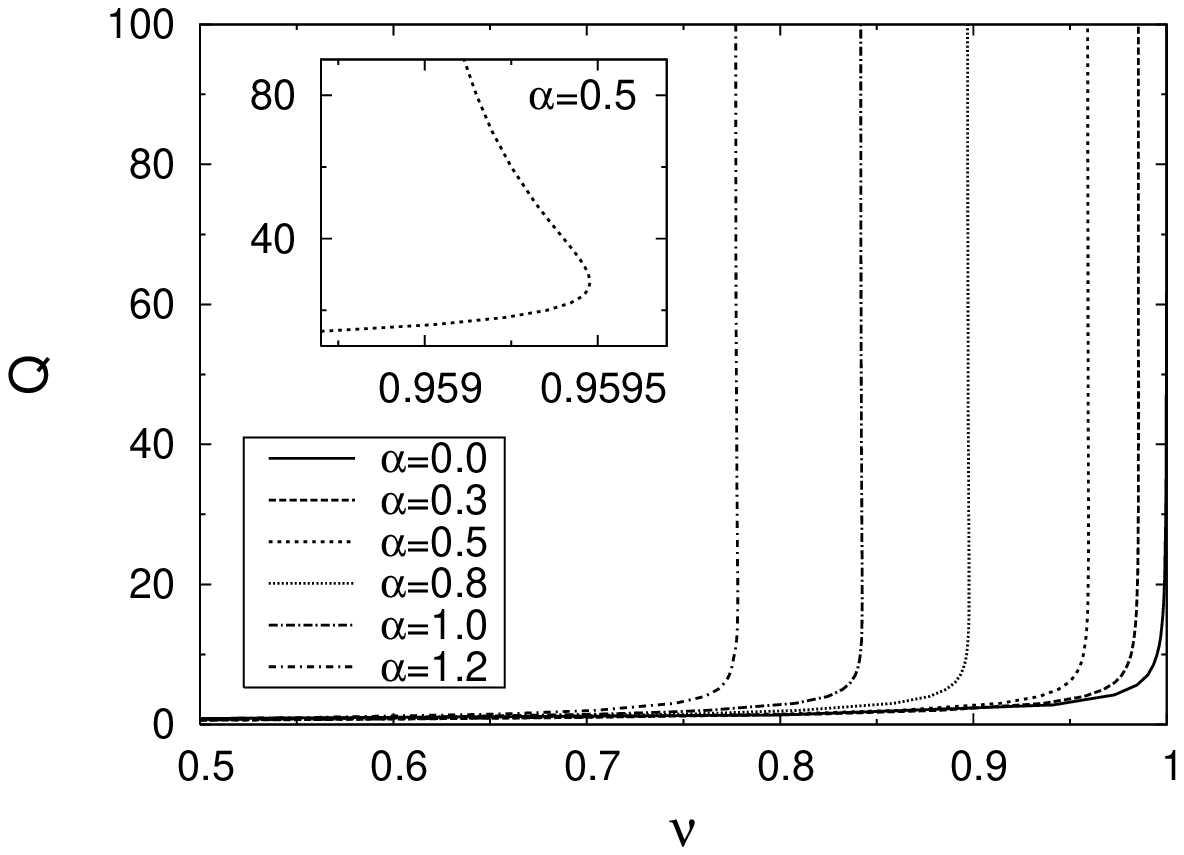}
\hspace{0.5cm} (b)\hspace{-0.3cm}
\includegraphics[height=.25\textheight, angle =0]{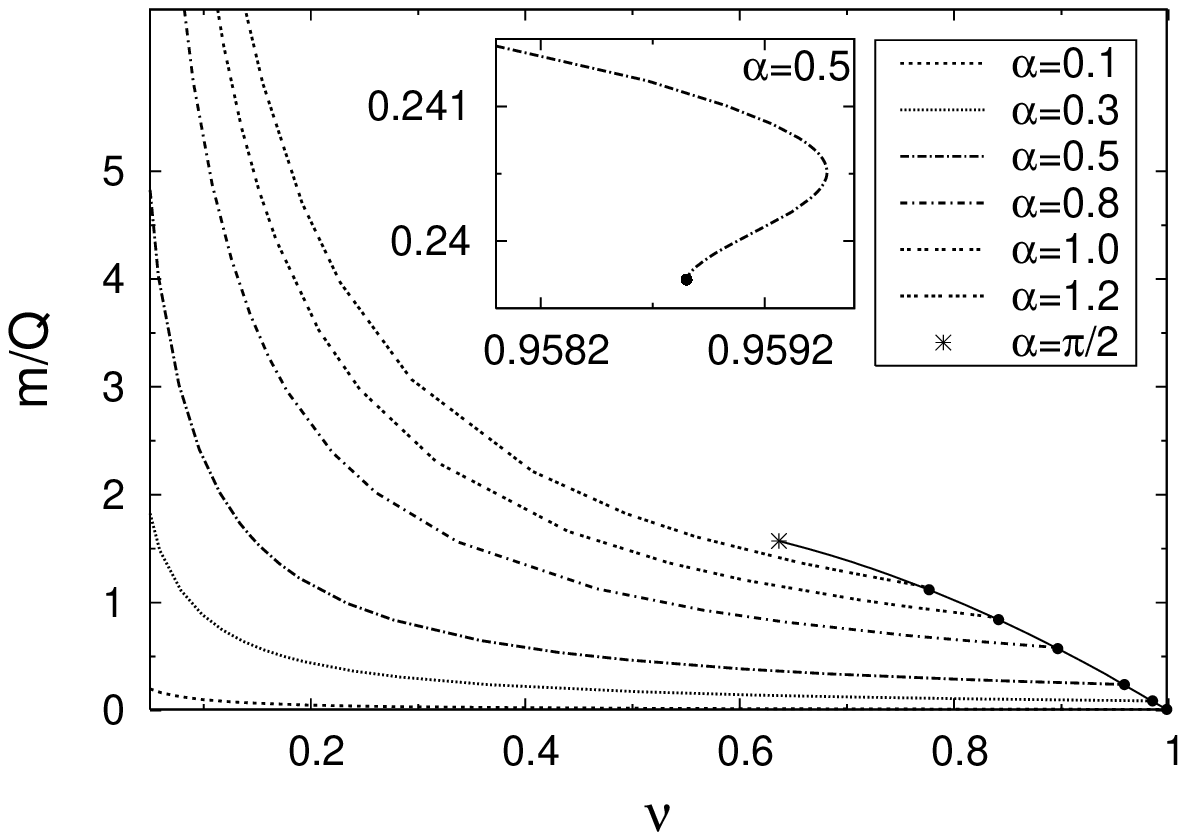}
% free line does it
\end{center}
\vspace{-0.5cm}
\caption{\small
 Dyons:
 (a) electric charge $Q(\nu)$ versus $\nu$,
 (b) scaled mass $\bar m=m/Q$ versus $\nu$,
     for several values of $\alpha$.
%\vspace{0.5cm}
}
\end{figure}

As seen in Fig.~\ref{f-1},
the second branches rise very steeply as $Q$ becomes large.
They are confined
to the intervals $\nu_{\rm cr}(\alpha) \le \nu \le \nu_{\rm max}(\alpha)$.
The lower bounds $\nu_{\rm cr}(\alpha)$,
approached in the limit of infinite charge,
are very close to the upper bounds $\nu_{\rm max}(\alpha)$.
The critical values $\nu_{\rm cr}(\alpha)$
are exhibited in Fig.~\ref{f-2}.
We note, that for small $\alpha$,
the critical values $\nu_{\rm cr}(\alpha)$
exhibit an almost quadratic dependence on $\alpha$.
No solutions are found beyond $\alpha=\pi/2$.

Let us now consider dyons for very large values of $Q$,
in order to identify the solution obtained
in the limit $Q \rightarrow \infty$.
Clearly, the mass $m$ diverges in the limit $Q \rightarrow \infty$.
However, the scaled mass $\bar m=m/Q$ tends to a finite limiting value,
which depends on the coupling strength $\alpha$,
as seen in Fig.~\ref{f-1}.
Another quantity of interest is the scalar charge
$c_{\rm H}$, determining the $1/r$ power law decay of the Higgs field
of the dyon.
Like the scaled mass,
the scaled scalar charge $\bar c_{\rm H} = c_{\rm H}/Q$ 
approaches a finite limiting value,
depending on the coupling strength $\alpha$.
In Fig.~\ref{f-2}
$\bar c_{\rm H}$ is shown versus $\alpha$
for a very large value of the electric charge, $Q=10000$.
%Clearly, the dependence on $\alpha$ looks like a simple
%cosine relation, $c_{\rm H}= \cos(\alpha)$.

\begin{figure}[h!]
\lbfig{f-2}
\begin{center}
\hspace{-0.5cm} (a)\hspace{-0.3cm}
\includegraphics[height=.25\textheight, angle =0]{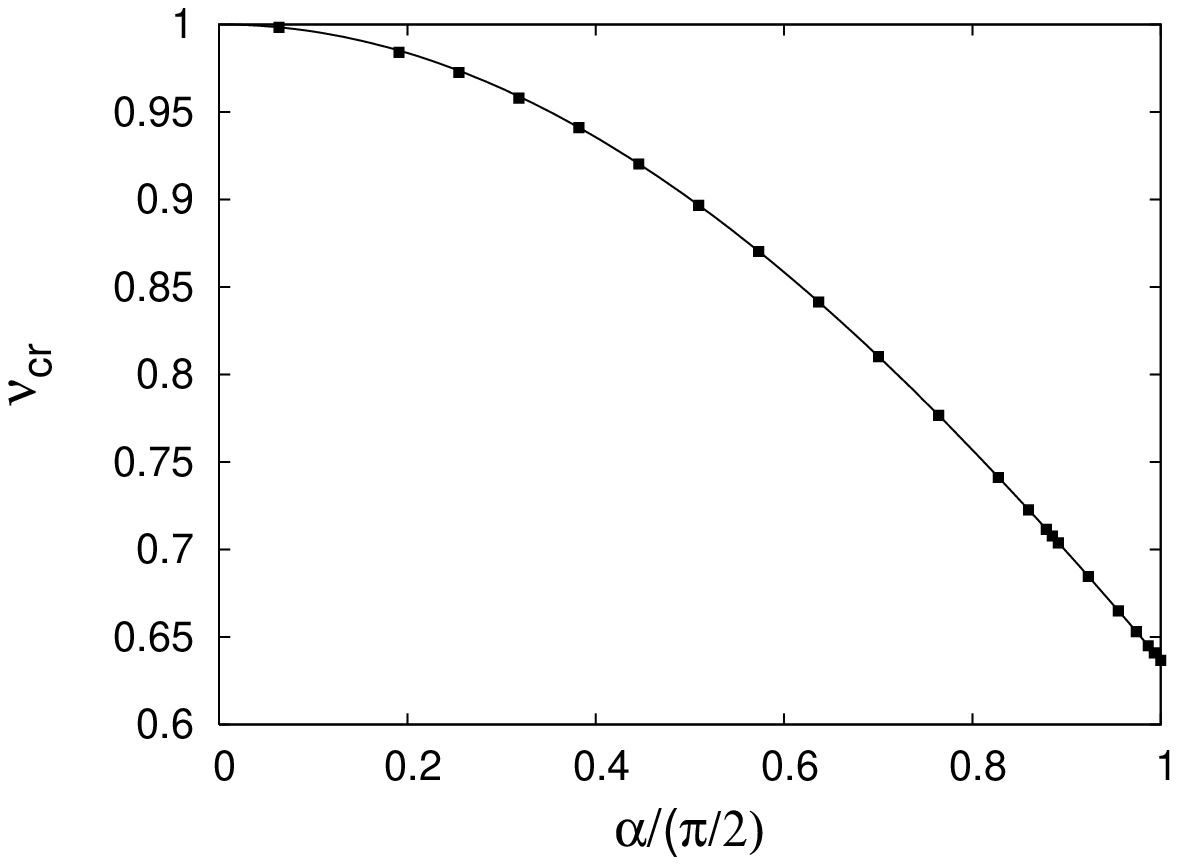}
\hspace{0.5cm} (b)\hspace{-0.3cm}
\includegraphics[height=.25\textheight, angle =0]{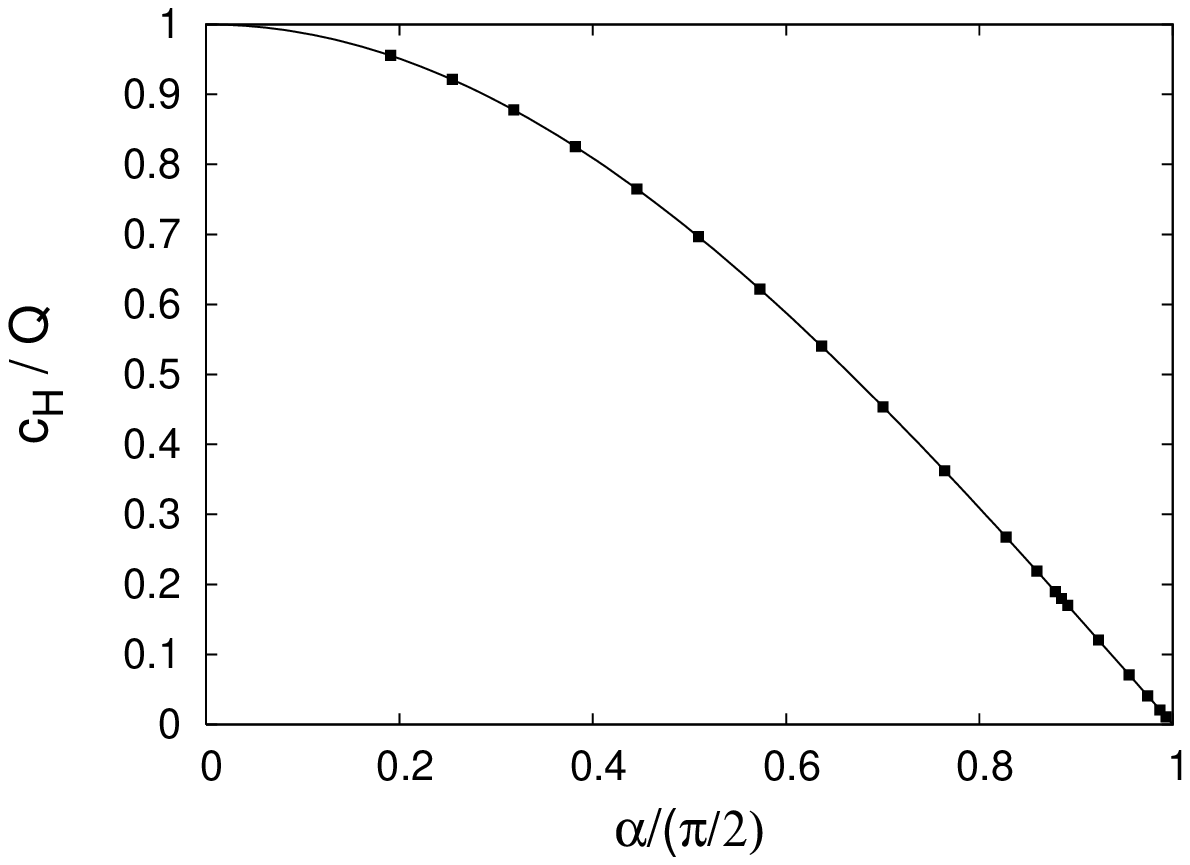}
\end{center}
\vspace{-0.5cm}
\caption{\small
Dyons:
 (a) critical values $\nu_{\rm cr}$ versus $\alpha$,
% compared to an approximate quadratic function
 (b) scaled scalar charge $\bar c_{\rm H} = c_{\rm H}/Q$ 
     versus $\alpha$.
 The exact dependence as obtained from the Penney solution
 is also exhibited for $\nu_{\rm cr}$ and $\bar c_{\rm H}$.
}
\end{figure}

\subsection{Relation with the Penney solutions}

To understand the limiting behaviour,
we now consider the solutions themselves 
for very large values of $Q$.
We exhibit in Fig.~\ref{f-3} the metric and matter functions
in scaled Schwarzschild-like coordinates,
$\bar r= r/Q$, for
a dyon solution with very large electric charge, $Q=10000$,
at a coupling strength $\alpha=0.5$.
The solution then appears to consist of two parts,
an interior part in the region $0 \le \bar r \le \bar r_0$,
and an exterior part in the region $\bar r_0 \le \bar r \le \infty$,
with $\bar r_0=1$.

\begin{figure}[h!]
\lbfig{f-3}
\begin{center}
\hspace{-0.5cm} (a)\hspace{-0.3cm}
\includegraphics[height=.25\textheight, angle =0]{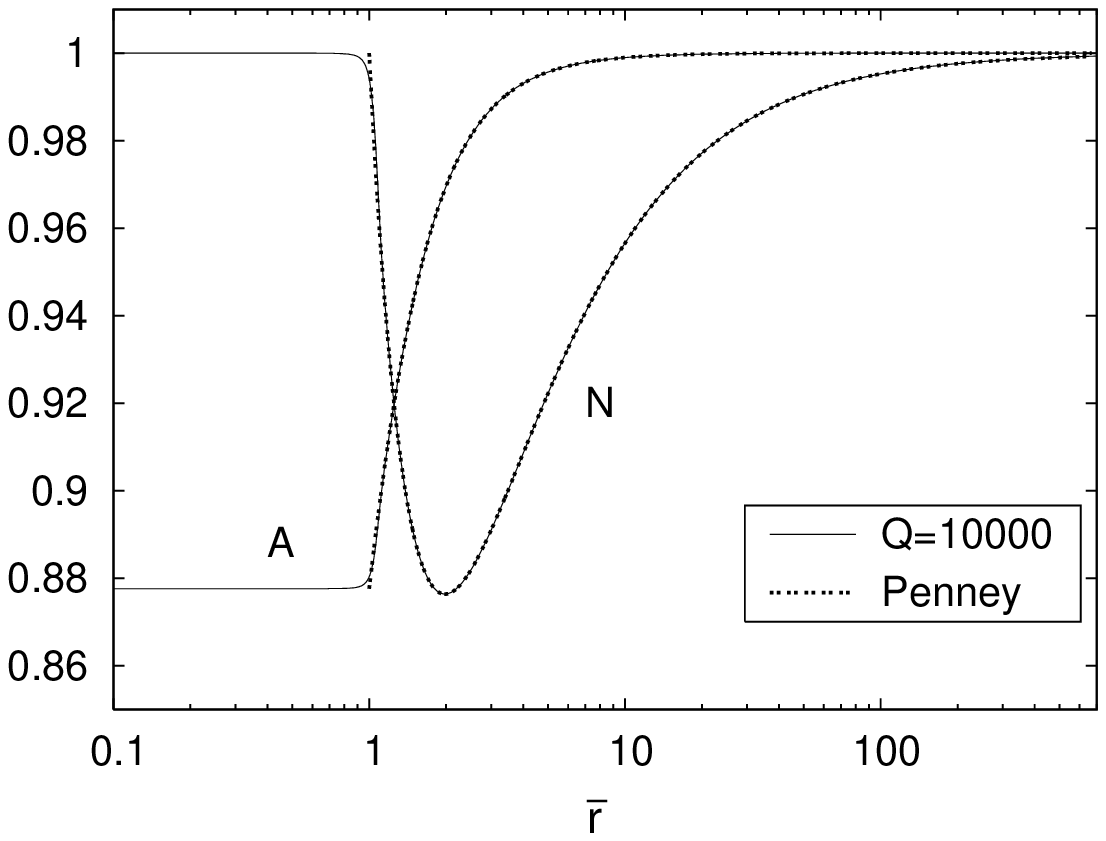}
\hspace{0.5cm} (b)\hspace{-0.3cm}
\includegraphics[height=.25\textheight, angle =0]{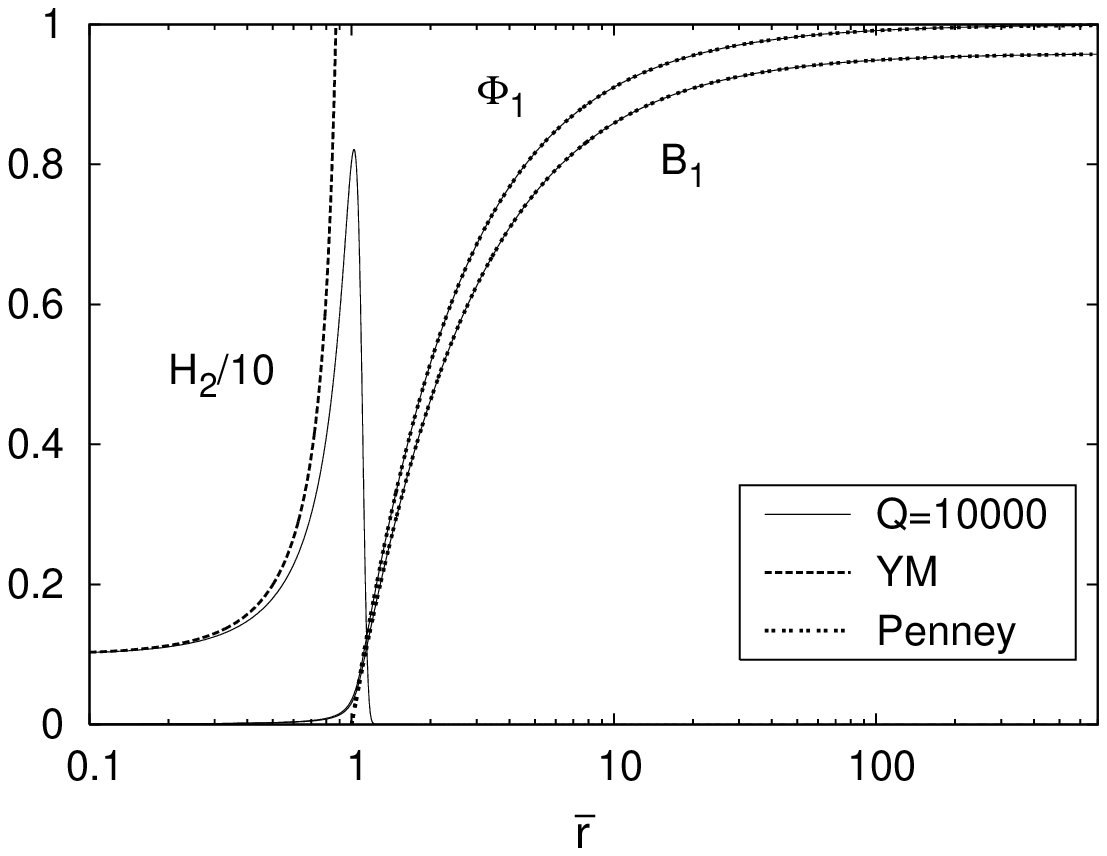}
\end{center}
\vspace{-0.5cm}
\caption{\small
Dyons:
  (a) metric functions $A$ and $N$,
  (b) gauge potential and Higgs field functions $\hat B_1$, $H_2$, $\Phi_1$,
      versus scaled Schwarzschild-like coordinate $\bar r=r/Q$,
      for an almost limiting solution at $\alpha=0.5$, $Q=10000$.
      Also shown is the limiting solution:
      the Penney solution in the exterior, and the flat non-Abelian
      solution in the interior.
}
\end{figure}

In the interior region, $0 \le \bar r \le 1$,
the limiting solution is given by
\begin{equation}
A(\bar r) =  {\rm const} \ , \ \ \
N(\bar r) = 1 \ , \ \ \
\hat B_1(\bar r)=\Phi_1(\bar r)=0 \ , \ \ \ H_2=H_2(\bar r) \ ,
\end{equation}
i.e., the metric is flat
and the time-component of the gauge potential and the Higgs field
both vanish. 
The only non-trivial function is the 
spatial gauge potential function $H_2$,
satisfying the single decoupled equation, Eq.~(\ref{eqH2}).

In the exterior region $1 \le \bar r \le \infty$, on the other hand,
the spatial gauge potential function $H_2$ vanishes,
while the metric and the other two matter functions satisfy
the coupled set of equations Eqs.~(\ref{eqmu})-(\ref{eqP1}).
This system represents a special case of
the coupled Einstein-Maxwell-scalar equations,
where the scaled electric charge has the value $\bar Q=1$,
%(because we scaled with $Q$).
studied in a different parametrization by Penney \cite{Penney:1969xk}.

%The Einstein-Maxwell-scalar equations were
%studied in a different parametrization by Penney \cite{Penney:1969xk}.
%We review the formulae of the Penney solution
%in the Appendix, focussing on the case
%relevant for the present discussion, 
%where in the Reissner-Nordstr\"om limit
%no horizons occur, but a naked singularity is present.

As discussed in section III, in the relevant Penney solutions
no horizons occur, but a naked singularity is present.
%We note that the relevant Penney solution % (with $g_{tt} > 0$)
In the limit $Q \rightarrow \infty$, these Penney solutions
precisely saturate the bound on the scalar charge, Eq.~(\ref{rel3}), 
i.e., they satisfy a quadratic relation between the scaled global charges,
\begin{equation}
\bar m^2 = \alpha^2 \left( 1 - \bar c_{\rm H}^2 \right) \ .
\label{massrelation2} \end{equation}
Due to the dyon boundary conditions
for the Maxwell and scalar functions the Penney solutions
%yields upon integration the relations Eq.~(\ref{rel8}) and Eq.~(\ref{rel9}),
furthermore satisfy the relations Eq.~(\ref{rel8}) and Eq.~(\ref{rel9}),
%determining 
i.e., 
the scaled scalar charge $\bar c_{\rm H}$,
and the potential parameter $\nu$ are given
in terms of the coupling strength $\alpha$,
\begin{equation}
\bar c_{\rm H}= \cos \alpha \ , \ \ \
\nu = \frac{\sin \alpha}{\alpha} \ ,
\label{cH} \end{equation}
and thus the scaled mass satisfies
$\bar m = \alpha \sin \alpha$.
The validity of these relations for the
limiting dyon solutions is seen in Fig.~\ref{f-1}
for the scaled mass $\bar m =m/Q$ 
and in Fig.~\ref{f-2} for $\nu$ and $\bar c_{\rm H} = c_{\rm H}/Q$.

Identifying the transition point $\bar r_0$
as the outer extremum of the metric functions $g_{tt}$ and $g_{rr}$
of this particular Penney solution,
we find $\bar r_0=1$ (Eq.~(\ref{rel10})).
Evaluating the metric functions $A$ and $N$ at 
the transition point $\bar r_0=1$ yields
\begin{equation}
A(\bar r_0)=  \bar c_{\rm H} \ , \ \ \ N(\bar r_0)=1 \ .
\label{bcpenney} \end{equation}

For comparison we superimpose in Fig.~\ref{f-3}
the limiting solutions in the two regions.
In the exterior region $\bar r_0 \le \bar r \le \infty$.
the limiting solution is the Penney solution 
with the same $\alpha$ and $\bar Q=1$,
saturating the mass bound,
Eq.~(\ref{massrelation2}).
This Penney solution 
also determines the constant metric functions
in the interior region $0 \le \bar r \le \bar r_0$,
via their boundary values at the transition point $\bar r_0$,
Eq.~(\ref{bcpenney}).
On the other hand, the boundary conditions at the origin 
determine the constant matter functions $\hat B_1$ and $\Phi_1$
in the interior,
and thus also at the transition point $\bar r_0$.

In contrast the matter function $H_2$ is a solution of
Eq.~(\ref{eqH2}) \cite{Rosen}. % \cite{foot3}. % and diverges at $\bar r_0$???
Expanding the solution $H_2(\bar{r})$
of the Yang-Mills equation Eq.~(\ref{eqH2}) in a power series,
$H_2(\bar{r}) = 1 +h_1 \bar{r} + h_2\bar{r}^2/2 +O(\bar{r}^3)$,
shows that $h_1=0$ and $h_2$ is a free parameter, characterizing the
solution.
Solving the equation numerically in the interval $0\leq \bar{r}\leq 1$
with boundary conditions $H_2(0)=1$, $H_2(1)=h^\ast$,
and varying $h^\ast$,
we observe that $h_2(h^\ast)$ increases with
increasing $h^\ast$ and tends to a finite value $h_2^{\max}\approx 3.047$
when $h^\ast$ tends to infinity.
Comparison with the numerical dyon solutions for large charge $Q$,
we find some evidence that $H _2^{''}(0)$ indeed tends to the value
$h_2^{\max}$ for $Q \rightarrow \infty$,
i.e., $H_2$ diverges at $\bar r=1$ in the limit.
Convergence %$H _2^{''}(0) \rightarrow h_2^{\max}$ is
is very slow, however, approximately like $1/\sqrt{Q}$.

We finally consider the range of $\alpha$ where
dyon solutions exist.
For small electric charge it was shown before \cite{gdyon},
that dyons exist only below a maximal value of $\alpha$,
e.g., $\alpha_{\rm max}(Q=0)=1.40$ and $\alpha_{\rm max}(Q=1)=1.41$.
As the charge is increased further, $\alpha_{\rm max}$ increases as well.
In the limit $Q \rightarrow \infty$,
we obtain from the above considerations % , Eq.~(\ref{cH}), 
a bound for $\alpha$,
\begin{equation}
0 \le \alpha \le \frac{\pi}{2} \ . \label{alphabound}
\end{equation}

Concerning the limit $\alpha \rightarrow \pi/2$ we see 
that the scaled scalar charge $\bar c_{\rm H}$ tends to zero.
Thus in the limit there is no (non-trivial) scalar field in the exterior,
whereas there is still an electric field.
This suggests, that for $\alpha \rightarrow \pi/2$
the scaled limiting solution corresponds in the exterior
to an extremal Reissner-Nordstr\"om solution
with charge $Q=\alpha$.
Inspection of the analytical formulae
and the numerical solutions shows,
that this is indeed the case.
We exhibit the approach towards this limit for the
metric and matter functions in Fig.~\ref{f-4}.
We note, however, that in this limit, the RN solution
is approached only in the interval $\pi/2 \le \bar r < \infty$.
The limiting solution is discontinuous at $\pi/2$,
since the metric function $A(\bar r)$ 
assumes the RN value $A=1$ only for $ \bar r > \pi/2$,
whereas it vanishes for $ \bar r < \pi/2$,
causing the metric to differ from the RN metric in this region.

\begin{figure}[t!]
\lbfig{f-4}
\begin{center}
\hspace{-0.5cm} (a)\hspace{-0.3cm}
\includegraphics[height=.25\textheight, angle =0]{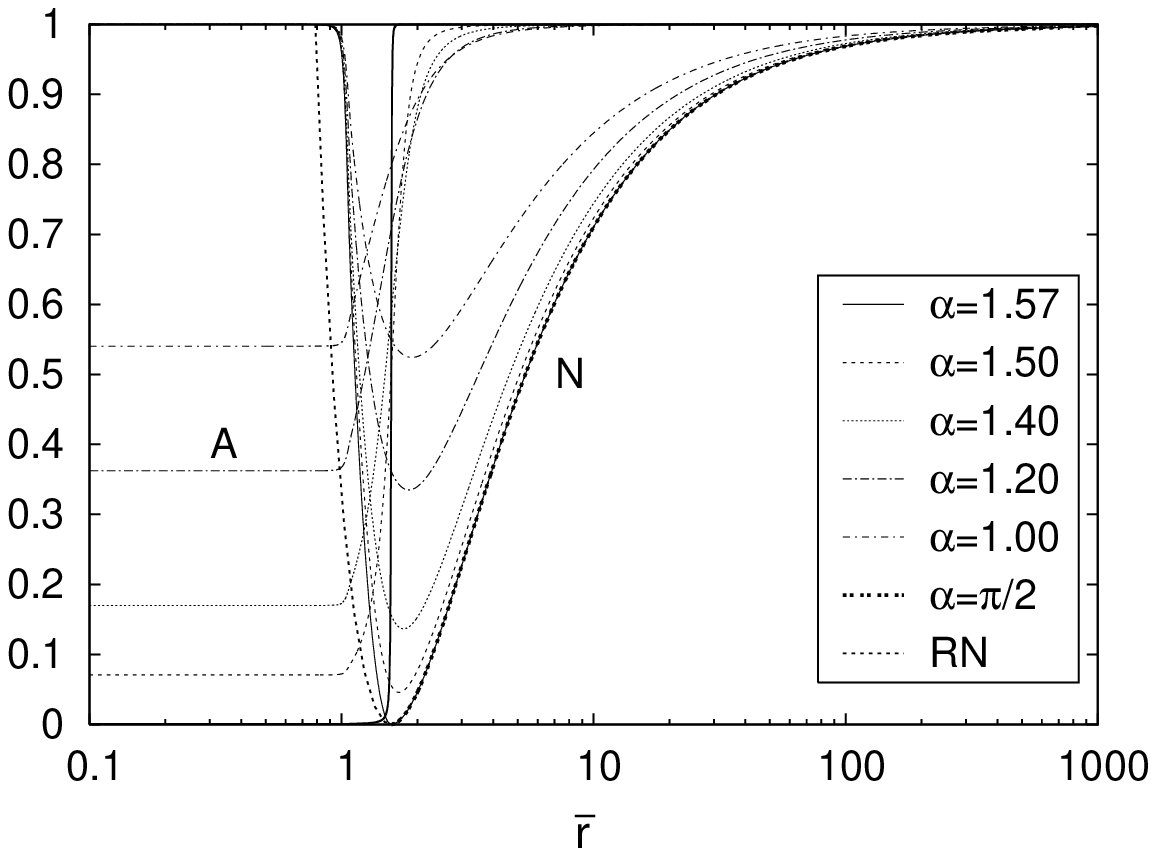}
\hspace{0.5cm} (b)\hspace{-0.3cm}
\includegraphics[height=.25\textheight, angle =0]{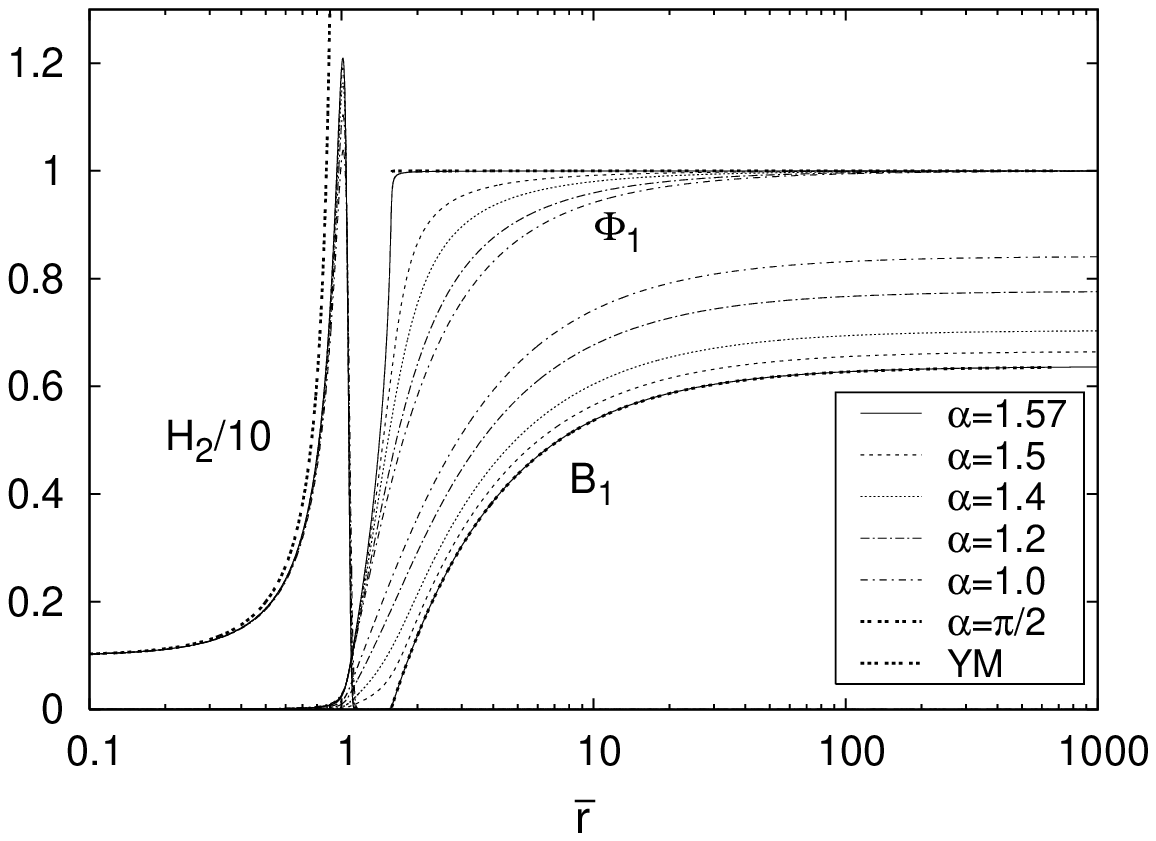}
\end{center}
\vspace{-0.5cm}
\caption{\small
Dyons:
  (a) metric functions $A$ and $N$,
  (b) gauge potential and Higgs field functions $\hat B_1$, $H_2$, $\Phi_1$,
      versus scaled Schwarzschild-like coordinate $\bar r=r/Q$,
      for solutions with $Q=10000$ and several values of $\alpha$,
      approaching $\alpha=\pi/2$.
      Also shown is the limiting solution:
      the extremal RN solution in the exterior, and the flat non-Abelian
      solution in the interior.
}
\end{figure}

Concluding,
we see that the scaled limiting solution consists of two parts: 
an Abelian exterior Penney solution,
saturating the bound Eq.~(\ref{massrelation2})
and determined by the value of $\alpha$ (with $\bar Q=1$),
and a non-Abelian interior part with flat metric,
where the constant metric functions are determined by the exterior
Penney solution at the transition point $\bar r_0=1$.

\section{Conclusions}

We have reconsidered dyon solutions of Einstein-Yang-Mills-Higgs theory,
in the limit of vanishing Higgs potential.
Dyon solutions then exist for arbitrarily large values of the charge.
As the charge becomes large, 
the coupled system of Einstein-Yang-Mills-Higgs equations
decomposes into a single equation for the magnetic gauge potential,
and a coupled system of equations for the metric, the electric
gauge potential and the scalar field.
This system of equations is equivalent to the set of
Einstein-Maxwell-scalar equations, studied by Penney
in a different parametrization
\cite{Penney:1969xk}.

In the limit $Q \rightarrow \infty$,
the non-Abelian dyons solutions then
tend to limiting solutions, which consist of two parts,
%(after scaling w.r.t.~$Q$)
a non-Abelian interior solution with flat metric,
and a gravitating Abelian exterior part.
The interior solutions represent 
non-trivial solutions of the single
equation for the magnetic gauge potential
(with the remaining equations trivially satisfied).
They are regular at the origin, but diverge at the transition point,
where the parts are joined.
%All other functions are constant, and thus the metric is flat.
The exterior solutions, on the other hand,
correspond to the exterior part of particular Penney solutions,
which then determine the asymptotic properties of the solutions,
such as their mass and their scalar charge.
%parameter $\nu$, the scalar charge,
%and the mass of the limiting solutions.

It appears interesting to also consider
monopole-antimonopole pairs with large electric charge.
When electric charge is added to a monopole-antimonopole pair,
the monopole and antimonopole experience a repulsive force
and the poles move further apart \cite{dyonhkk}.
More importantly, however, the pair begins to rotate about its symmetry axis
with an angular momentum $J$, equal to the product of
the total electric charge $Q$ and the
(magnitude of the) individual magnetic charge $n$
of the constituent magnetic poles,
$J=nQ$ \cite{radu}.

Preliminary study shows, that the presence of electric charge
also leads to bifurcations of the 
charged monopole-antimonopole solutions,
with new branches of solutions arising.
Considering larger values of the charge, we observe,
that the values of $\nu$, $\bar c_H$ and $\bar m$
appear to be consistent with the respective Penney relations.
This seems surprising though, since the Penney solutions
carry no angular momentum, while the charged dyons
possess angular momentum $J=nQ$.
%However, due to numerical difficulties,
%the calculations currently cannot be extended
%to very large values of the charge,
%as would be needed.

\begin{acknowledgments}
We thank Andrey Shoom for bringing the Penney solution
to our attention.
R.I. gratefully acknowledges support by the Volkswagenstiftung,
and B.K. support by the German Aerospace Center.
\end{acknowledgments}


\begin{thebibliography}{000}

\bibitem{mono}G.~`t Hooft, 
              Nucl.\ Phys.\ {\bf B79} (1974) 276;\\
              A.~M. Polyakov, 
              Pis'ma JETP {\bf 20} (1974) 430.

\bibitem{WeinbergGuth}
                      E.J.~Weinberg, and A.H.~Guth,
                      Phys. Rev. {\bf D14} (1976) 1660.

\bibitem{RebbiRossi}
                    C.~Rebbi, and P.~Rossi,
                    Phys. Rev. {\bf D22} (1980) 2010.

\bibitem{mmono}
               R.S.~Ward,
               Comm. Math. Phys. {\bf 79} (1981) 317;\\
               P.~Forgacs, Z.~Horvath, and L.~Palla,
               Phys. Lett. {\bf 99B} (1981) 232;\\
               M.K.~Prasad,
               Comm. Math. Phys. {\bf 80} (1981) 137;\\
               M.K.~Prasad, and P.~Rossi,
               Phys. Rev. {\bf D24} (1981) 2182.

\bibitem{KKT} 
              B. Kleihaus, J. Kunz, and D.~H. Tchrakian,
              Mod. Phys. Lett. {\bf A13} (1998) 2523.

\bibitem{monoDS}
 N.~J. Hitchin, N.~S. Manton and M.~K. Murray,
 %Symmetric monopoles,
 Nonlinearity {\bf 8} (1995) 661;\\
 C.~J. Houghton and P.~M. Sutcliffe,
 %Tetrahedral and cubic monopoles,
 Commun. Math. Phys. {\bf 180} (1996)343;\\
 C.~J. Houghton and P.~M. Sutcliffe,
 %Octahedral and dodecahedral monopoles,
 Nonlinearity {\bf 9} (1996) 385;\\
 P.~M. Sutcliffe,
 Int. J. Mod. Phys. {\bf A 12} (1997) 4663;\\
 C.~J. Houghton, N.~S. Manton and P.~M. Sutcliffe,
 Nucl. Phys. {\bf B 510} (1998) 507.

\bibitem{dyon}
 B. Julia and A. Zee,
%Poles with both magnetic and electric charges in non-abelian gauge theory,
 Phys. Rev. {\bf D11} (1975) 2227;\\
%\bibitem{PraSom}
 M.~K.~Prasad, and C.~M.~Sommerfeld,
 Phys. Rev. Lett. {\bf 35} (1975) 760.

\bibitem{dyonhkk}
 B.~Hartmann, B.~Kleihaus, and J.~Kunz,
 Mod. Phys. Lett. A {\bf 15} (2000) 1003.

\bibitem{HSV}
 M. Heusler, N. Straumann, and M. Volkov,
 Phys. Rev. {\bf D58} (1998) 105021.

\bibitem{radu}
 J.~J. van der Bij and E. Radu,
%On rotating regular nonabelian solutions,
Int. J. Mod. Phys. {\bf A17} (2002) 1477;\\
Int. J. Mod. Phys. {\bf A18} (2003) 2379.

\bibitem{Taubes}
 C.~H. Taubes,
 Commun. Math. Phys. {\bf 86} (1982) 257; \\
 C.~H. Taubes,
 Commun. Math. Phys. {\bf 86} (1982) 299;\\
 C.~H. Taubes,
 Commun. Math. Phys. {\bf 97} (1985) 473.

\bibitem{Rueber}
 W. Nahm, unpublished;\\
 B. R\"uber, Thesis, University of Bonn 1985.

\bibitem{mapKK}
 B.~Kleihaus, and J.~Kunz,
 Phys. Rev. {\bf D61} (2000) 025003.

\bibitem{KKS}
              B.~Kleihaus, J.~Kunz, and Ya.~Shnir,
              %Monopole--Antimonopole Chains,
              Phys. Lett. {\bf B570}, (2003) 237;\\
              B.~Kleihaus, J.~Kunz, and Ya.~Shnir,
              %Monopoles, Antimonopoles and Vortex Rings,
              Phys. Rev. {\bf D68} (2003) 101701(R);\\
              B.~Kleihaus, J.~Kunz, and Ya.~Shnir,
              %Monopole--Antimonopole Chains and Vortex Rings,
              Phys. Rev. {\bf D70} (2004) 065010.

\bibitem{KNS}
J. Kunz, U. Neemann, and Ya. Shnir,
%Transitions between Vortex Rings and Monopole--Antimonopole Chains,
%e-Print: hep-th/0606176
Phys. Lett. {\bf B640} (2006) 57.

\bibitem{gmono}
 K.~Lee, V.P.~Nair, and E.J.~Weinberg,
 Phys. Rev. {\bf D45} (1992) 2751;\\
 P.~Breitenlohner, P.~Forgacs, and D.~Maison,
 Nucl. Phys. {\bf B383} (1992) 357;\\
 P.~Breitenlohner, P.~Forgacs, and D.~Maison,
 Nucl. Phys. {\bf B442} (1995) 126.

\bibitem{HKK}
 B.~Hartmann, B.~Kleihaus, and J.~Kunz,
 Phys. Rev. Lett. {\bf 86} (2001) 1422;\\
 B.~Hartmann, B.~Kleihaus, and J.~Kunz,
 Phys. Rev. {\bf D65} (2001) 024027.

\bibitem{gdyon}
 Y. Brihaye, B. Hartmann, and J. Kunz,
 Phys. Lett. {\bf B441} (1998) 77;\\
 Y. Brihaye, B. Hartmann, J. Kunz, and N. Tell,
 Phys. Rev. {\bf D60} (1999) 104016.

\bibitem{MAP}
 B.~Kleihaus, and J.~Kunz,  
 Phys. Rev. Lett. {\bf 85} (2000) 2430. 

\bibitem{PRT}
 V. Paturyan, E. Radu, and D.~H. Tchrakian,
%Rotating regular solutions in Einstein-Yang-Mills-Higgs theory,
 Phys. Lett. {\bf B609} (2005) 360.

\bibitem{KKN}
 B. Kleihaus, J. Kunz, and U. Neemann,
%Gravitating Stationary Dyons and Rotating Vortex Rings,
 Phys. Lett. {\bf B623} (2005) 171.

\bibitem{KKNN}
 B. Kleihaus, J. Kunz, F. Navarro-L\'erida,and U. Neemann,
%Stationary Dyonic Regular and Black Hole Solutions,
 e-Print: arXiv:0705.1511 [gr-qc].

\bibitem{KKS4}
 B.~Kleihaus, J.~Kunz, and Ya.~Shnir,
%Gravitating Monopole--Antimonopole Chains and Vortex Rings,
 Phys. Rev. {\bf D71} (2005) 024013;\\
 J. Kunz, U. Neemann, and Ya. Shnir,
%Gravitating Monopole-Antimonopole Systems at Large Scalar Coupling,
 Phys. Rev. {\bf D75} (2007) 125008.

\bibitem{BM}
 R.~Bartnik, and J.~McKinnon,  
 Phys. Rev. Lett. {\bf 61} (1988) 141.

\bibitem{KK}
 B.~Kleihaus, and J.~Kunz,  
 Phys. Rev. Lett. {\bf 78} (1997) 2527;\\
 B.~Kleihaus, and J.~Kunz,  
 Phys. Rev. {\bf D57} (1998) 834.

\bibitem{IKKS}
 R. Ibadov, B. Kleihaus, J. Kunz, and Ya. Shnir,
 Phys. Lett. {\bf B609} (2005) 150.
 
\bibitem{Kari}
%\cite{Kleihaus:2005pg}
%\bibitem{Kleihaus:2005pg}
  B.~Kleihaus, J.~Kunz and K.~Myklevoll,
  %``Excited platonic sphalerons in the presence of a dilaton field,''
  Phys.\ Lett.\  B {\bf 632}, 333 (2006)
  [arXiv:hep-th/0509106].
  %%CITATION = PHLTA,B632,333;%%

%\cite{Penney:1969xk}
\bibitem{Penney:1969xk}
  R.~Penney,
  %``Generalization of the reissner-nordstroem solution to the einstein field
  %equations,''
  Phys.\ Rev.\  {\bf 182}, 1383 (1969).
  %%CITATION = PHRVA,182,1383;%%
  
%\cite{Rosen:1972uu}
\bibitem{Rosen}
  G.~Rosen,
  %``Exact solutions to the yang-mills field equations,''
  J.\ Math.\ Phys.\  {\bf 13} (1972) 595;
  %%CITATION = JMAPA,13,595;%%
  see also 
  A.~Actor,
  %``Classical Solutions Of SU(2) Yang-Mills Theories,''
  Rev.\ Mod.\ Phys.\  {\bf 51} (1979) 461.
  %%CITATION = RMPHA,51,461;%%

  

  

\end{thebibliography}
\end{document}